\documentclass[journal]{IEEEtran}

\usepackage{graphicx}
\usepackage{amsmath,amssymb}
\usepackage{booktabs}
\usepackage[hidelinks]{hyperref}
\usepackage{siunitx}

\begin{document}

\title{An LLM-Orchestrated Agent for Directional-Coupler Design with
Self-Consistent Eigenmode and FDTD Validation}


\author{
    \IEEEauthorblockN{Saumya Biswas, Amrit De, and Md Tauhidul Islam}%
    \thanks{Manuscript submitted \today. Saumya Biswas was with the Department of
    Mechanical Engineering, University of Maryland, College Park, MD, USA
    (e-mail: biswas.saumya@yahoo.com).  Corresponding author: Saumya Biswas.}%
    \thanks{Amrit De is with Apsidal LLC, Riverside, California, USA (e-mail: ade@apsidal.org).}%
    \thanks{Md Tauhidul Islam is with the Department of Radiation Oncology, Stanford University, United States (e-mail: tauhid@stanford.edu).}%
}


\markboth{IEEE Journal of Quantum Electronics}%
{Author \MakeLowercase{\textit{et al.}}: LLM-Orchestrated Directional-Coupler Design}

\maketitle

\begin{abstract}
We present a design agent which is a Large Language Model (LLM) that orchestrates,
but does not perform, the numerical simulations to design a silicon-on-insulator (SOI)
$2\times2$ directional coupler. We choose a symmetric phase-matched coupler where a lot of analytical results are available that help the design strategy. The LLM proposes candidate gap values (a geometrical dimension size) and judges convergence, while all physics is owned by deterministic solvers: a
frequency-domain eigenmode solver estimates the coupling coefficient~$\kappa$ for the current design, and an independent Finite-Difference Time-Domain (FDTD) stage validates it. Both solvers operate on a common slab-projected two-dimensional (2D)
effective-index reduction of the silicon film, so the design~$\kappa$ and the FDTD
response are consistent by problem design; the residual between them is shown to be
a single constant phase offset~$\phi$, attributable to a fixed excess coupling
length $L_{\mathrm{extra}}=\SI{2.837(11)}{\micro\meter}$ that we find invariant
across a factor-of-two range in~$\kappa$. Folding this offset into a closed-loop
length correction, the agent delivers a $50/50$ splitter whose FDTD-measured cross
fraction is $0.498$ (target $0.500$), a residual of $0.0017$. Results are
made self-consistent within the 2D effective-index model; and the LLM succeeds in delivering a suitable design over a number of attempts.
\end{abstract}


\section{Introduction}
\IEEEPARstart{C}{ertain} photonic component design requires intelligent design iterations by an experienced engineer. Currently, the field of numerical algorithms and softwares for photonic design is mature with industry-scale development and deployment, especially in the industrially important Silicon photonics \cite{chrostowski2019silicon,bogaerts2018silicon,photonics2004silicon}. Design automation has always been an important objective in the field \cite{heins2016design,mingaleev2015towards}. The advent of Large Language Models (LLM) has significantly enhanced the prospects of design automation \cite{chiarello2024generative,pan2025survey,liu2024toward}. LLM's functionality is not limited to performing analysis tasks or offering judgments with passive text generation, it can possibly serve as an autonomous agent accomplishing some task \cite{yao2022react}, especially using tools \cite{schick2023toolformer}.

A fundamental design component of large scale photonic circuits is the Directional Coupler (DC). Its design is grounded in well established physical theories of electromagnetism, the coupled-mode theory \cite{haus1991coupled} in particular. The coupled mode theory has a long legacy of successful application in the specific problem of waveguide physics \cite{Yariv1973,Huang1994,Liu2005,Okamoto2006}. The specific problem this work is focused on is the simultaneous tuning of the waveguide gap and interaction length
of two waveguides so that input light is split between two output ports in a prescribed ratio. The mechanism is well understood in terms of classical electromagnetics. In the conventional approach to the problem, the parameter space is scanned quickly through a fast approximate model and verified with a slower, higher-fidelity solver. Harmonizing the two models is the engineering task, and ad-hoc automation processes may now be sought through LLMs that automate the interactions between the two methods. 

We are not interested in enabling the LLM to perform sophisticated physical simulation, but rather investigate efficient agent architecture for performing a task. We therefore narrow down our attention to a symmetric phase-matched directional coupler where a lot of analytical results help reduce the complexity of the physics part of the problem \cite{ElSaeed2024}. For the particular problem, the supermode splitting uniquely determines the coupling coefficient. We fixate on a 2D effective-index model, and the coupling coefficient is obtainable from the even-odd supermode index splitting \cite{ElSaeed2024},
\begin{eqnarray}
\kappa = \pi \left( n_{eff, even} - n_{eff, odd} \right)/ \lambda, \ \ \ \ \label{eq_kappa}
\end{eqnarray}
where $n_{eff}$ is the effective refractive index of the supermodes. A frequency domain eigenmode solver (MPB) \cite{Johnson2001} calculates the guided supermodes and extracts the coupling coefficient $\kappa$ with Eq. \eqref{eq_kappa}. We obtain the desired interaction length necessary for the target power-splitting ratio. The resulting design is independently verified with MEEP \cite{Oskooi2010}--a Finte difference Time Domain (FDTD) solver furnishing the transmitted and cross-coupled power ratios. Beyond proposing gaps for the next iteration, the LLM is tasked with convergence supervision, autonomous recovery from failed proposals, and a supervisory control layer role overall.

In this work, We demonstrate an agent structure that can successfully orchestrate the iterative validation of the deterministic computation parts. The reasoning function of the LLM is separated from the deterministic computation tasks. The LLM proposes the next candidate geometry for evaluation, and judges the achievement of convergence. The calibration workflow incorporates the separate LLM judgement and trusted numerics routines consistently into a design annd validation framework. We base the electromagnetic problem on previously reported methods of dimensional reduction \cite{knox1970integrated}, the effective index method
(EIM) and correct for similar effects observed in other works \cite{nikkhah2024inverse}. The MPB is an eigenmode calculation of a 2D slab and in FDTD calculation, we pack the height dimension into an effective index--making the iterations much faster. A constant phase correction is required which relates to a per-waveguide wavenumber offset observed elsewhere \cite{nikkhah2024inverse}, but is characteristically different here (not waveguide-number dependent). The combined closed-loop correction (for phase) leads to the agent delivering the target split to within $0.2\%$. The reasoning of the LLM agent is successful for the specific geometry and physics.

\section{Device Geometry and the 2D Model}
\label{sec:geometry}
In the The device under study is a symmetric 2$\times$2 DC. Two parallel
silicon strip waveguides on a Silicon-On-Insulator (SOI) platform constitutes the DC. The width of each waveguide is $w=\SI{500}{\nano\meter}$ and height is $h=\SI{220}{\nano\meter}$. The silicon index $n_{\mathrm{Si}}=3.476$ at the operating wavelength $\lambda=\SI{1550}{\nano\meter}$. The separation of the two cores i.e. the lateral gap $g$ is the design variable. It is swept over
\SIrange{160}{235}{\nano\meter}. The silica cladding has index
$n_{\mathrm{SiO_2}}=1.444$. We idealize the geometry with the assumption that buried oxide below (underneath the device film) and the cladding above have the same index (a fabricated device will not typically have this symmetry). 

A 3D FDTD simulation of the coupler would be computationally demanding since we would need the converged 3D run repeatedly (once for each agent loop). We opt for a 2D Effective-Tndex
Method: the out-of-plane confinement of the \SI{220}{\nano\meter} film is
traded for a single effective slab index $n_{\mathrm{eff,slab}}$. The reduced model is a completely in-plane problem in which each core appears as a region of index
$n_{\mathrm{eff,slab}}$ embedded in an oxide background. Consequently, the FDTD evaluations become fast enough to support the multi-length sweeps that the validation and
phase-correction stages require.

The slab index is the lumped parameter embodying the compaction of the vertical dimension. It is not a fitted or assumed free parameter, but rather the effective
index of the fundamental TE-polarized mode of the vertical stack--a \SI{220}{\nano\meter} silicon film surrounded by oxide claddings. It is obtained from the solution of the symmetric three-layer slab dispersion relation for the TE$_0$ mode (at $\lambda=\SI{1550}{\nano\meter}$),
$
\tan\!\left(\frac{\kappa_y h}{2}\right) = \frac{\gamma}{\kappa_y},
\quad
\kappa_y = k_0\sqrt{n_{\mathrm{Si}}^2 - n^2},
\quad
\gamma = k_0\sqrt{n^2 - n_{\mathrm{SiO_2}}^2}$ with $k_0=2\pi/\lambda$ and $n$ being the modal index \cite{Okamoto2006,YarivYeh2007}. We round the solution $n_{\mathrm{eff,slab}}=2.848$ to $2.85$. In the 2D reduction, the vertical evanescent tails above and below the film are subsumed into $n_{\mathrm{eff,slab}}$, so the
2D FDTD explicitly deals with only the lateral evanescent coupling across the gap. It is also revealing for the underlying physics since the design variable $g$ comes to the fore only in the physics of the in-plane overlap.

\section{Agent Architecture}
\begin{figure}[t]
\centering
\includegraphics[width=\columnwidth]{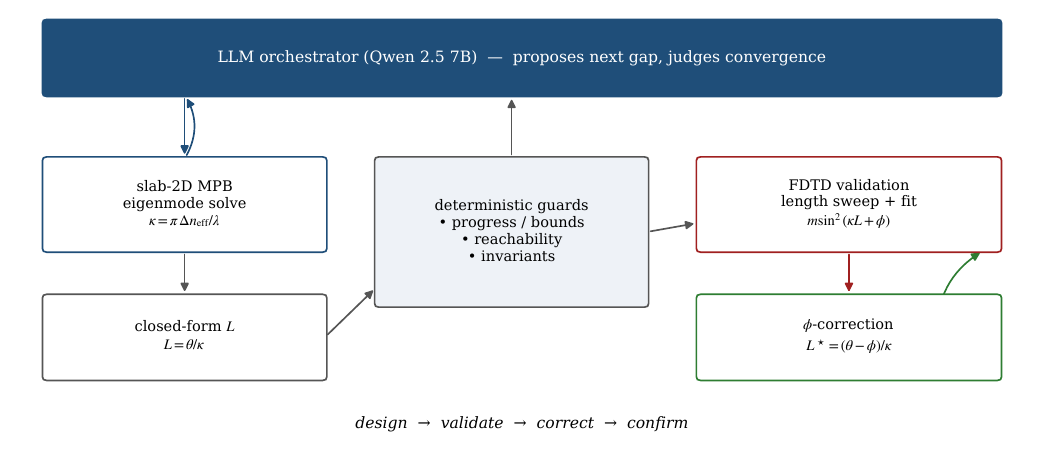}
\caption{Agent architecture. The LLM orchestrator proposes the next gap and
evaluates convergence; deterministic solvers and guards own all numerics; the physics is off-limit for the LLM. The
eigenmode solver supplies~$\kappa$ for design, an independent FDTD stage
validates by a length sweep and fit, and a closed-loop $\phi$-correction sets the
delivered length. The overall flow is design~$\rightarrow$ validate~$\rightarrow$
correct~$\rightarrow$ confirm.}
\label{fig:arch}
\end{figure}

The LLM (Qwen~2.5, 7B parameters) is utilized by the agent for the whole loop of agent workflow (top of Fig.~\ref{fig:arch}). The agent is tasked with leading the design work for a target split ratio and a gap search interval. With the frequency domain eigenmode (MPB) solver, the agent calculates the even and odd supermode indices and using Eq. \eqref{eq_kappa}, the coupling coefficient $\kappa$.

The value furnishes the values for the interaction length in closed form as
$L=\theta/\kappa$ with $\theta=\arcsin\sqrt{P^\star}$ (~\cite{Yariv1973,Huang1994,Liu2005,Okamoto2006}) for the target cross
fraction~$P^\star$. The agent uses deterministic guards for bound checks. Every new proposal must pass through the progress and reachibility test against the gap interval, and physical intervals. The LLM makes informed decisions from interacting with these gatekeepers. The LLM only serves as the mediator of these bound and plausibility checks and the numerical routines. LLM does not itself compute any physical simulation loop, and cannot propose a simulation result for any iteration. It's job is restricted to proposing geometrical attributes and judging convergence. A poor proposal may waste an iteration but
cannot corrupt fidelty of the final device being designed.

\section{Convergence Behavior}
\begin{figure}[t]
\centering
\includegraphics[width=\columnwidth]{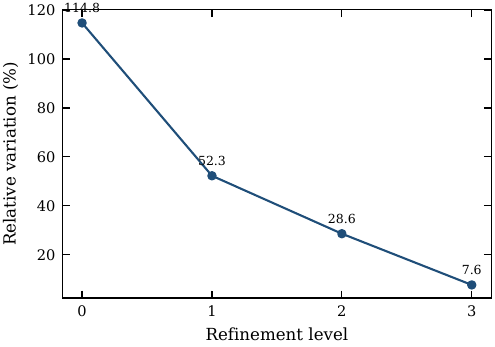}
\caption{Relative variation of the predicted split plotted against refinement level. The
quantity decreases monotonically ($114.8 \rightarrow 52.3 \rightarrow 28.6
\rightarrow 7.6\%$), indicating a stably converging solve.}
\label{fig:conv}
\end{figure}

The purpose of the agent design here is to mimic an enperienced engineer's strategies. The agent does not simply trust the simulation results from a given round, it sweeps the solver discretization and examines the relative variation of the predicted split. Fig.~\ref{fig:conv} shows this relative variations over refinement levels falling
monotonically from $114.8\%$ at the coarsest level to $7.6\%$ (after three
refinements). Thus, the agent also orchestrates a validation loop over discretizations to attest the convergence of the solution. The monotone decrease in Fig.~\ref{fig:conv} assures us that the solve is well-defined and that the quantity the agent judges the convergence of, is genuinely converging. Thus, the agent's search serves as a self-consistency check of the physics under study as well.

\section{Calibrating Design Against Validation}

\subsection{A shared effective-index reduction \label{intro_Nikah}}
\begin{figure}[t]
\centering
\includegraphics[width=\columnwidth]{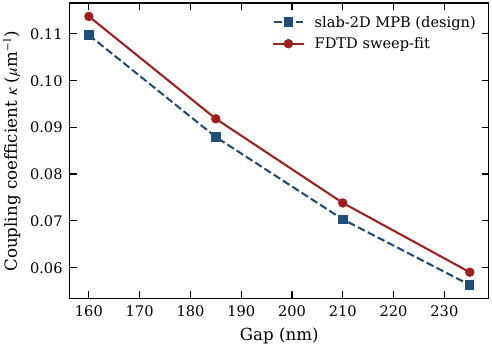}
\caption{Coupling coefficient~$\kappa$ plotted against gap from the slab-2D eigenmode model (design) and from the FDTD length-sweep fit (validation). In the effective index map, the two track each other consistently to within $\sim$$5\%$ across
the range; the small residual offset is the constant phase analyzed in
Section~\ref{sec:phi}.}
\label{fig:kappa}
\end{figure}

The DC design is built upon the collaboration between two disparate methods. The design success hinges upon the mutual consistency of the two methods. The theoretical consistency between the designer eigenmode model and verifier FDTD model dictates the final fidelity of the device designed. The two theories perform dissimilar calculations. One is an eigenmode calculation, the other a dynamics calculation. Still, the two methods share a reduced dimensional problem that is "effective index" matched. The cross-coupled power comes from the flux monitors with mode decomposition in the FDTD method. For this method the dynamical Maxwell's equations are solved on a grid. However, it can tap into the same physics as the frequency domain eigenmode solver when we consider the slab-projected description (sec~\ref{sec:geometry}). The vertical confinement of the
\SI{220}{\nano\meter} film is mapped onto the effective slab index
$n_{\mathrm{eff,slab}}=2.85$ (Eq.s in sec \ref{sec:geometry}). The $\kappa$ designed through supermode index separation and the FDTD response are built upon the same vertical resolution, and the physical theories converge on the same in-plane problem. Fig.~\ref{fig:kappa} shows a comparison of the slab-2D design~$\kappa$ with the FDTD-measured~$\kappa$ for a range of gaps from \SIrange{160}{235}{\nano\meter}. The two track each other closely, with the FDTD value keeping at a near-constant $\sim$$4$--$5\%$ value above the design value. 

The separation of the graphs in Fig.~\ref{fig:kappa} is not a numerical issue or a theoretical inconsistency. The fixed offset between two models that share the same slab
reduction is the consequence of a single constant phase. It is an expected feature of problems of the kind. In an unrelated inverse-design setting, Nikkhah
\emph{et al.}~\cite{nikkhah2024inverse} similarly find that their effective index reduction reproduces bulk propagation but inherits a (wave-number dependent) phase offset. Their effect can be compensated with a constant per-waveguide adjustment. We discuss our resolution (applied in the same spirit as~\cite{nikkhah2024inverse}) in the next subsection.

\subsection{Extracting \texorpdfstring{$\kappa$}{kappa} by a length-sweep fit}
\begin{figure}[t]
\centering
\includegraphics[width=\columnwidth]{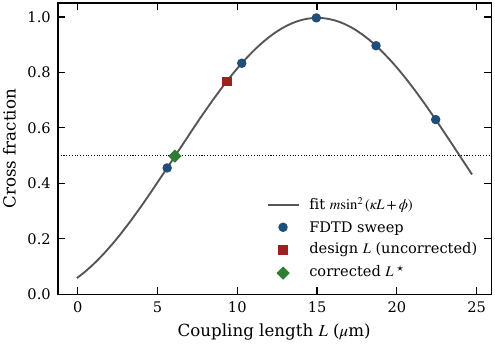}
\caption{Closed-loop result. FDTD-measured cross fraction versus coupling length (points) with the fitted $m\sin^2(\kappa L+\phi)$ model (line). The uncorrected
design length (square) overshoots $0.5$; the $\phi$-corrected length~$L^\star$
(diamond) lands on target at a measured cross fraction of $0.498$.}
\label{fig:closed}
\end{figure}

In this subsection, we give more details on the method for extracting $\kappa$ from the FDTD simulation. This turns out to be quite revealing, the underlying physics surfaces in the curve fitting that is done for the $\kappa$ extraction. The method requires sweeping the coupling length L, and fitting $P_{\mathrm{cross}}(L)$ with Eq. \eqref{eq:fit}. The FDTD~$\kappa$ is extracted from a sweep of coupling lengths. The FDTD is run at
several lengths bracketing the design point and a fit with Eq. \eqref{eq:fit} reveals the cross power trasnfer's dependence on the coupling strength as well as the phase offset. The equation
\begin{equation}
P_{\mathrm{cross}}(L) = m\,\sin^2(\kappa L + \phi),
\label{eq:fit}
\end{equation}
is adopted from the theory of bent couplers in~\cite{ElSaeed2024}. Throughout the range of interaction lengths (Fig.~\ref{fig:closed}), the fit returns $m \approx 0.99$ (a near-ideal symmetric coupler) and
$R^2 > 0.9999$ (the coefficient of determination), validating the functional form ~\eqref{eq:fit} that codifies the physics here.

\section{The Lead-In Phase and Its Constancy}
\label{sec:phi}
From the fact $\phi$ is consistently found to be a specific nonzero value across all fits, We can deduce that its origin is of geometric nature. The simulation cell contains two parallel waveguides that run the full span between the source and output monitors, so the interaction between waveguides acts over a distance somewhat longer than the nominal interaction length~$L$. These fixed margins on either side relates to the source and monitor placement, but not to the gap. The effective device
therefore behaves as a coupler of \emph{effective} length $L_{\mathrm{eff}} = L +
L_{\mathrm{extra}}$, and the excess length $L_{\mathrm{extra}}$ contributes the constant phase $\phi = \kappa L_{\mathrm{extra}}$. According to this theory, $L_{\mathrm{extra}}=\phi/\kappa$ should be a single length (independent of gap).

\begin{figure}[t]
\centering
\includegraphics[width=\columnwidth]{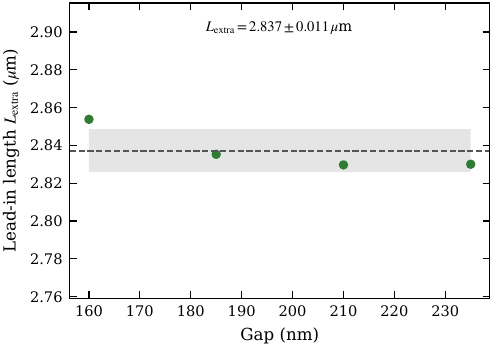}
\caption{Lead-in coupling length $L_{\mathrm{extra}}=\phi/\kappa$ plotted against the gap. The
value is more or less constant at $\SI{2.837(11)}{\micro\meter}$ (dashed line, shaded
$\pm1\sigma$) across a factor-of-two range in~$\kappa$, ascertaining that the phase
offset arises from a fixed lead-in geometry and is independent of the gap.}
\label{fig:lextra}
\end{figure}

Fig.~\ref{fig:lextra} clearly bears out these assumptions. Throughout the range of gaps spanning a factor-of-two
range in~$\kappa$, the extracted $L_{\mathrm{extra}}=\phi/\kappa$ is constant at
$\SI{2.837(11)}{\micro\meter}$, with a relative scatter of $0.40\%$. This gap
independence serves as the conclusive proof. The excess coupling length concomitant of the cell geometry is, by definition, the same at every coupling strength, whereas a phase
arising from a coupling-rate variation would scale with~$\kappa$ and would not collapse
onto a constant $L_{\mathrm{extra}}$. The offset is therefore not a range dependent parameter but a fixed, measurable property of the simulation geometry. Therefore, we have to estimate it only once rather than refit at every design configuration. Before closing out the section, we reiterate (from subsection \ref{intro_Nikah}) that effective-index reductions are known to produce residual phase offsets between reduced and higher-fidelity models. Nikkhah \emph{et al.}~\cite{nikkhah2024inverse}, rectify such an offset effect with
a constant per-waveguide adjustment. But, their offset arises from a slab-mode
wavenumber mismatch and, unlike our system, accumulates interface phase.

\section{Closed-Loop Length Correction}
In this section, we elucidate a division between the two roles of the agent. During the
gap search, the LLM proposes each candidate gap for the deterministic
guards to accept or reject. The LLMs judgement is required up to the achievement of the feasible operating point only. The remaining work: validating~$\kappa$, measuring~$\phi$, and correcting the length is executed entirely by the deterministic layer. The model does not participate in this part of the loop. The correction layer is merely a closed-form
computation, not a learned or model-mediated layer.

Eq.~\eqref{eq:fit} represents the true behaviour of the device, so a coupler built at the naive length
$L=\theta/\kappa$ overshoots the target by the phase~$\phi$ i.e. by the
excess coupling length $L_{\mathrm{extra}}=\phi/\kappa$ of Section~\ref{sec:phi}.
Starting with the validated $(\kappa,\phi)$ at the operating point, the deterministic layer computes the delivered straight-section length that achieves~$P^\star$ as
\begin{equation}
L^\star = \frac{\arcsin\sqrt{P^\star}-\phi}{\kappa},
\label{eq:correction}
\end{equation}
using the FDTD-fitted~$\kappa$ and~$\phi$ from the same validation sweep. As such, the correction is not limited by the residual design-versus-validation~$\kappa$
difference. Fig.~\ref{fig:closed} shows the final result for a $50/50$ target. The uncorrected design length lands at a cross fraction of $0.79$; which is very much off target. Post correction steps, the corrected length~$L^\star$ delivers an FDTD-measured cross fraction of
$0.498$ against the target $0.500$. The residual of $0.0017$ is at the level of the
FDTD discretization itself. This completes the design$\rightarrow$validate%
$\rightarrow$correct$\rightarrow$confirm loop, with the LLMs role as the proposer of gaps and deterministic layer correcting every reported quantity.

\section{Termination Behavior}
The process terminates in one out of four possible ways, and which one activates is decided by the
deterministic layer in every case. Three falls under the jurisdiction of the code: the residual-based \emph{convergence} test is met at the shortest feasible length;
the iteration count trigger by reaching its \emph{depth cap}, and the guard trips---an oscillation or overshoot in the residual halts the search. The fourth,
\emph{LLM-declined}, the important one, is the one path where the language model can end the loop. It may do so by proposing that no further gap is worth trying. Even then, the model only
proposes a stop, and the deterministic finalization decides what to return. This fine print matters: the model never certifies a design as converged. The judgment rests on the residual condition checked in code, so a premature or
over-eager model cannot mark an inadequate design as finished.

Table~\ref{tab:terminations} presents some representative runs. In the cases shown,
the search reached the depth cap and returned the best feasible design via
fallback, or the model declined to continue. In both cases the deterministic
finalization produced a usable design rather than a failure, which is the desired functionality of the guard layer.

\begin{table}[t]
\centering
\caption{Observed termination paths across agent runs.}
\label{tab:terminations}
\begin{tabular}{lcccc}
\hline
Termination & Converged & Fell back & Steps & LLM calls \\
\hline
depth cap & no & yes & 10 & 11 \\
llm declined & no & no & 4 & 8 \\
\hline
\end{tabular}
\end{table}

\section{Discussion and Limitations}
Our results demonstrate a modular approach to the problem of a DC design. The burdens are separated modularly--- LLM judgement above, and trusted numerical sofware below. The calibration code entwines the design and validation stages in a consistent framework. Successful executions (centrally by the closed-loop correction) deliver designs of $\kappa$ to within $0.2\%$. At the current stage of the work, the LLM is not given the power of final verdict, only deterministic judgement loops are entrusted with that responsibility. but, Table~\ref{tab:terminations} records the statistics of enough successful runs to justify the current agent architecture.

The central limitation is obvious and deliberate. Every result here lives within
a 2D effective-index model: the slab-2D eigenmode solve and the 2D FDTD treat the same physics by conscious design, and the validating simulation uses the same 2D model against which the correction was fit. The reported $0.498$ delivered split is not a prediction of a fabricated
three-dimensional device, but rather a statement of self-consistency within that model. The true coupling would depend on the full vertical structure that neither solver treats here. Extending the loop to a 3D eigenmode computation and 3D FDTD validation and characterizing how well $L_{\mathrm{extra}}$ and the
$\kappa$ agreement carry over---is the natural future direction and the prerequisite for any fabrication-accuracy claim. Next, $L_{\mathrm{extra}}$ is specific to the present source/monitor geometry and would require re-measurement if that geometry changed.

\section{Conclusion}
We have demonstrated a directional-coupler design agent that is a large language model. It orchestrates a loop of trusted deterministic solvers without computing any physics itself. Matching the dimensionality of the design and validation models with effective index strategies, we built a methodical loop attack for the problem. Interesting theoretical concepts are also observed and corrected for. Post mapping, the residual is a single constant lead-in phase, $\phi=\kappa L_{\mathrm{extra}}$ with $L_{\mathrm{extra}}=\SI{2.837(11)}{\micro\meter}$. Meaningfully, we found it invariant across the design range; and a closed-loop correction using the validated $(\kappa,\phi)$ delivers a $50/50$ split to within $0.0017$ in cross transfer power fraction. All results are
self-consistent within a 2D effective-index model, and the extension to
fabrication-accurate 3D prediction is identified as the natural future work. While in this work, the LLM only proposes new values for one geometrical measure, the gap, in future works, more variables may be considered.

\section*{Code and Data Availability}
The source code (the eigenmode and FDTD solver wrappers, the analysis notebooks, and the result files underlying every figure in this paper) are openly available at \url{https://github.com/sbisw002/Photonics-Agent-}. The repository
also includes the recorded run outputs, so the manuscript figures can be regenerated without re-executing the solvers, together with the standalone slab-mode script that derives the effective slab index $n_{\mathrm{eff,slab}}$ used throughout. Reproducing the full design loop additionally requires the open-source MPB and MEEP packages and a local Qwen~2.5 model served through Ollama, as documented in the repository.

\section*{Author Contributions}
[Saumya Biswas]: Conceptualization; Methodology; Software; Validation;
Formal analysis; Visualization; Writing -- original draft.
[Amrit De] Conceptualization of the photonic architecture problem.],
[Md Tauhidul Islam] Conceptualization of the Agent architecture.
All authors reviewed and approved the final manuscript.

\section*{Acknowledgment}
The authors used Anthropic's Claude Opus large language model to assist with code generation during the development of the design agent and its analysis scripts. All generated code was reviewed, tested, and validated by the authors, who take full responsibility for the content of this work.

\bibliographystyle{IEEEtran}
\bibliography{references}

@article{Oskooi2010,
  author  = {Oskooi, Ardavan F. and Roundy, David and Ibanescu, Mihai and
             Bermel, Peter and Joannopoulos, J. D. and Johnson, Steven G.},
  title   = {{MEEP}: A flexible free-software package for electromagnetic
             simulations by the {FDTD} method},
  journal = {Computer Physics Communications},
  volume  = {181},
  number  = {3},
  pages   = {687--702},
  year    = {2010},
  doi     = {10.1016/j.cpc.2009.11.008}
}

@inproceedings{knox1970integrated,
  title={Integrated circuits for the millimeter through optical frequency range},
  author={Knox, RM and Toulios, PP},
  booktitle={Proc. Symp. Submillimeter Waves},
  volume={20},
  pages={497--515},
  year={1970},
  organization={Brooklyn, NY}
}

@article{nikkhah2024inverse,
  title={Inverse-designed low-index-contrast structures on a silicon photonics platform for vector--matrix multiplication},
  author={Nikkhah, Vahid and Pirmoradi, Ali and Ashtiani, Farshid and Edwards, Brian and Aflatouni, Firooz and Engheta, Nader},
  journal={Nature Photonics},
  volume={18},
  number={5},
  pages={501--508},
  year={2024},
  publisher={Nature Publishing Group UK London}
}

@article{liu2024toward,
  title={Toward automated simulation research workflow through LLM prompt engineering design},
  author={Liu, Zhihan and Chai, Yubo and Li, Jianfeng},
  journal={Journal of Chemical Information and Modeling},
  volume={65},
  number={1},
  pages={114--124},
  year={2024},
  publisher={ACS Publications}
}

@article{pan2025survey,
  title={A survey of research in large language models for electronic design automation},
  author={Pan, Jingyu and Zhou, Guanglei and Chang, Chen-Chia and Jacobson, Isaac and Hu, Jiang and Chen, Yiran},
  journal={ACM Transactions on Design Automation of Electronic Systems},
  volume={30},
  number={3},
  pages={1--21},
  year={2025},
  publisher={ACM New York, NY}
}

@article{chiarello2024generative,
  title={Generative large language models in engineering design: opportunities and challenges},
  author={Chiarello, Filippo and Barandoni, Simone and {\v{S}}kec, Marija Majda and Fantoni, Gualtiero},
  journal={Proceedings of the Design Society},
  volume={4},
  pages={1959--1968},
  year={2024},
  publisher={Cambridge University Press}
}

@inproceedings{mingaleev2015towards,
  title={Towards an automated design framework for large-scale photonic integrated circuits},
  author={Mingaleev, Sergei and Richter, Andr{\'e} and Sokolov, Eugene and Arellano, Cristina and Koltchanov, Igor},
  booktitle={Integrated Optics: Physics and Simulations II},
  volume={9516},
  pages={951602},
  year={2015},
  organization={SPIE}
}

@incollection{heins2016design,
  title={Design flow automation for silicon photonics: Challenges, collaboration, and standardization},
  author={Heins, Mitchell and Cone, Chris and Ferguson, John and Cao, Ruping and Pond, James and Klein, Jackson and Korthorst, Twan and Bakker, Arjen and Stoffer, Remco and Fiers, Martin and others},
  booktitle={Silicon Photonics III: Systems and Applications},
  pages={99--156},
  year={2016},
  publisher={Springer}
}

@article{photonics2004silicon,
  title={Silicon photonics},
  author={Photonics, Silicon},
  journal={Topics in Applied Physics},
  volume={94},
  year={2004}
}

@article{bogaerts2018silicon,
  title={Silicon photonics circuit design: methods, tools and challenges},
  author={Bogaerts, Wim and Chrostowski, Lukas},
  journal={Laser \& Photonics Reviews},
  volume={12},
  number={4},
  pages={1700237},
  year={2018},
  publisher={Wiley Online Library}
}

@article{chrostowski2019silicon,
  title={Silicon photonic circuit design using rapid prototyping foundry process design kits},
  author={Chrostowski, Lukas and Shoman, Hossam and Hammood, Mustafa and Yun, Han and Jhoja, Jaspreet and Luan, Enxiao and Lin, Stephen and Mistry, Ajay and Witt, Donald and Jaeger, Nicolas AF and others},
  journal={IEEE Journal of Selected Topics in Quantum Electronics},
  volume={25},
  number={5},
  pages={1--26},
  year={2019},
  publisher={IEEE}
}

@article{schick2023toolformer,
  title={Toolformer: Language models can teach themselves to use tools},
  author={Schick, Timo and Dwivedi-Yu, Jane and Dess{\`\i}, Roberto and Raileanu, Roberta and Lomeli, Maria and Hambro, Eric and Zettlemoyer, Luke and Cancedda, Nicola and Scialom, Thomas},
  journal={Advances in neural information processing systems},
  volume={36},
  pages={68539--68551},
  year={2023}
}

@article{yao2022react,
  title={React: Synergizing reasoning and acting in language models},
  author={Yao, Shunyu and Zhao, Jeffrey and Yu, Dian and Du, Nan and Shafran, Izhak and Narasimhan, Karthik and Cao, Yuan},
  journal={arXiv preprint arXiv:2210.03629},
  year={2022}
}

@article{haus1991coupled,
  title={Coupled-mode theory},
  author={Haus, Hermann A and Huang, Weiping},
  journal={Proceedings of the IEEE},
  volume={79},
  number={10},
  pages={1505--1518},
  year={1991},
  publisher={IEEE}
}

@article{Johnson2001,
  author  = {Johnson, Steven G. and Joannopoulos, J. D.},
  title   = {Block-iterative frequency-domain methods for {Maxwell's}
             equations in a planewave basis},
  journal = {Optics Express},
  volume  = {8},
  number  = {3},
  pages   = {173--190},
  year    = {2001},
  doi     = {10.1364/OE.8.000173}
}

@article{Yariv1973,
  author  = {Yariv, Amnon},
  title   = {Coupled-mode theory for guided-wave optics},
  journal = {IEEE Journal of Quantum Electronics},
  volume  = {9},
  number  = {9},
  pages   = {919--933},
  year    = {1973},
  doi     = {10.1109/JQE.1973.1077767}
}

@book{Liu2005,
  author    = {Liu, Jia-Ming},
  title     = {Photonic Devices},
  publisher = {Cambridge University Press},
  address   = {Cambridge, UK},
  year      = {2005}
}

@book{Okamoto2006,
  author    = {Okamoto, Katsunari},
  title     = {Fundamentals of Optical Waveguides},
  edition   = {2},
  publisher = {Academic Press},
  address   = {Burlington, MA},
  year      = {2006}
}

@article{Huang1994,
  author  = {Huang, Wei-Ping},
  title   = {Coupled-mode theory for optical waveguides: an overview},
  journal = {Journal of the Optical Society of America A},
  volume  = {11},
  number  = {3},
  pages   = {963--983},
  year    = {1994},
  doi     = {10.1364/JOSAA.11.000963}
}

@book{YarivYeh2007,
  author    = {Yariv, Amnon and Yeh, Pochi},
  title     = {Photonics: Optical Electronics in Modern Communications},
  edition   = {6},
  publisher = {Oxford University Press},
  address   = {New York},
  year      = {2007}
}

@article{ElSaeed2024,
  author  = {El-Saeed, Ahmed H. and Elshazly, Alaa and Kobbi, Hakim and
             Magdziak, Rafal and Lepage, Guy and Marchese, Chiara and
             Rahimi Vaskasi, Javad and Bipul, Swetanshu and Bode, Dieter and
             Ersek Filipcic, Marko and Velenis, Dimitrios and
             Chakrabarti, Maumita and De Heyn, Peter and Verheyen, Peter and
             Absil, Philippe and Ferraro, Filippo and Ban, Yoojin and
             Van Campenhout, Joris and Bogaerts, Wim and Deng, Qingzhong},
  title   = {Low-Loss Silicon Directional Coupler with Arbitrary Coupling
             Ratios for Broadband Wavelength Operation Based on Bent
             Waveguides},
  journal = {arXiv preprint arXiv:2404.06117},
  year    = {2024},
  eprint  = {2404.06117},
  archivePrefix = {arXiv},
  primaryClass  = {physics.optics}
}

\end{document}